\documentstyle[twoside,fleqn,espcrc2,epsf]{article}
\begin{document}


\begin{flushright}
LPTENS-97/43 \\
{\tt hep-th@xxx/9710136}\\
October 1997
\end{flushright}
\begin{center}
\vspace{3 ex}
{\Large\bf M(atrix) theory : a pedagogical introduction $^{*}$}\\
\vspace{8 ex}
Adel Bilal\thanks{This work is partially
supported by the European Commision under TMR contract
ERBFMRX-CT96-0045.} \\
{\it CNRS - Laboratoire de Physique Th\'eorique de l'\'Ecole
        Normale Sup\'erieure}\\
{\it 24 rue Lhomond, 75231 Paris Cedex 05, France} \\
 {\tt bilal@physique.ens.fr}\\
\vspace{15 ex}
\bf Abstract\\
\end{center}
\vspace{2 ex}
I attempt to give a pedagogical introduction to the matrix model of M-theory
as developed by Banks, Fischler, Shenker and Susskind (BFSS). In the first
lecture, I introduce and review the relevant aspects of D-branes with the
emergence of the matrix model action. The second lecture deals with the
appearance of eleven-dimensional supergravity and M-theory in strongly coupled
type IIA superstring theory. The third lecture combines the  materiel of the
two previous ones to arrive at the BFSS conjecture and explains the evidence
presented by these authors. The emphasis is not on most recent developments 
but on a hopefully pedagogical presentation.
\vspace{15 ex}

\centerline{{\it Lectures given at the Ecole Normale Sup\'erieure, Paris
and at the conference}}
\centerline{{\it ``Quantum aspects of gauge theories,
supersymmetry and unification", Neuch\^atel, September 1997}}
 
\vskip 3.cm
\hrule 
\vskip 2.mm
\noindent
$^{*}$ Work partially
supported by the European Commision under TMR contract
ERBFMRX-CT96-0045.

\thispagestyle{empty}
\newpage

\newcommand{\fig}[3]{\epsfxsize=#1\epsfysize=#2\epsfbox{#3}}

\def\PL #1 #2 #3 {Phys.~Lett.~{\bf #1} (#2) #3}
\def\NP #1 #2 #3 {Nucl.~Phys.~{\bf #1} (#2) #3}
 \def\PR #1 #2 #3 {Phys.~Rev.~{\bf #1} (#2) #3}
 \def\PRL #1 #2 #3 {Phys.~Rev.~Lett.~{\bf #1} (#2) #3}
 \def\CMP #1 #2 #3 {Comm.~Math.~Phys.~{\bf #1} (#2) #3}
 \def\IJMP #1 #2 #3 {Int.~J.~Mod.~Phys.~{\bf #1} (#2) #3}
 \def\JETP #1 #2 #3 {Sov.~Phys.~JETP.~{\bf #1} (#2) #3}
 \def\PRS #1 #2 #3 {Proc.~Roy.~Soc.~{\bf #1} (#2) #3}
 \def\IM #1 #2 #3 {Inv.~Math.~{\bf #1} (#2) #3}
 \def\JFA #1 #2 #3 {J.~Funkt.~Anal.~{\bf #1} (#2) #3}
 \def\LMP #1 #2 #3 {Lett.~Math.~Phys.~{\bf #1} (#2) #3}
 \def\IJMP #1 #2 #3 {Int.~J.~Mod.~Phys.~{\bf #1} (#2) #3}
 \def\FAA #1 #2 #3 {Funct.~Anal.~Appl.~{\bf #1} (#2) #3}
 \def\AP #1 #2 #3 {Ann.~Phys.~{\bf #1} (#2) #3}
 \def\MPL #1 #2 #3 {Mod.~Phys.~Lett.~{\bf #1} (#2) #3}

\def\d{\partial}
\def\db{\overline\partial}
\def\dt{\partial_t}
\def\zb{{\overline z}}
\def\f{\phi}
\def\vf{\varphi}
\def\th{\theta}
\def\itime{\int {\rm d} t \, }
\def\e{{\rm e}}
\def\a{\alpha}
\def\b{\beta}
\def\o{\omega}
\def\m{\mu}
\def\n{\nu}
\def\r{\rho}
\def\l{\lambda}
\def\g{\gamma}
\def\pe{p_{11}}
\def\rmd{{\rm d}}
\def\rd{\sqrt{2}}
\def\la{\langle}
\def\ra{\rangle}
\def\vac{\vert 0\rangle}
\def\mod{{\rm mod}}
\def\P{\Psi}
\def\p{\psi}
\def\dd #1 #2{{\p #1\over \p #2}}
\def\tr{{\rm tr}\, }
\def\Tr{{\rm Tr}\, }
\def\til{\widetilde}
\def\N{\nabla}
\def\b{\beta}
\def\s{\sigma}
\def\t{\tau}

\hyphenation{Suss-kind}

\title{M(atrix) theory : a pedagogical introduction $^{*}$}

\author{A. Bilal\address
{CNRS - Laboratoire de Physique Th\'eorique de l'Ecole
Normale Sup\'erieure, \\
 24 rue Lhomond, 75231
Paris Cedex 05, France,\\
{e-mail:} {\tt bilal@physique.ens.fr}
}}


\begin{abstract}
I attempt to give a pedagogical introduction to the matrix model of M-theory
as developed by Banks, Fischler, Shenker and Susskind (BFSS). In the first
lecture, I introduce and review the relevant aspects of D-branes with the
emergence of the matrix model action. The second lecture deals with the
appearance of eleven-dimensional supergravity and M-theory in strongly coupled
type IIA superstring theory. The third lecture combines the  materiel of the
two previous ones to arrive at the BFSS conjecture and explains the evidence
presented by these authors. The emphasis is not on most recent developments 
but on a hopefully pedagogical presentation.
\end{abstract}

\maketitle

\section{INTRODUCTION}

Among the multitude of dramatic developments in string duality during the
last three years maybe the  the most striking one has been the return of
eleven dimensional supergravity. The strong-coupling limit of the low-energy
sector of type IIA superstring is eleven dimensional supergravity. Since
eleven dimensional supergravity by itself does not seem to be a consistent
quantum theory, while the full superstring theory does, the question
immediately arose what is the consistent quantum theory that is the
strong-coupling limit of the full type IIA superstring - not only of its
low-energy sector. This theory was named M-theory, with many possible
interpretations for the letter ``M". Thus the two things we know about
M-theory are that it is the strong coupling limit of type IIA superstring and
that its low-energy limit is eleven dimensional supergravity. Though one was
lacking an intrinsic definition of M-theory in terms of its underlying degrees
of freedom, its mere existence led to many powerful predictions or
simplifications of superstring dualities.

A major step forward was taken by Banks, Fisch\-ler, Shenker and Susskind 
 \cite{BFSS} when they conjectured that the microscopic degrees of freedom of
M-theory when described in a certain Lorentz frame are D0-branes. The Lorentz
frame in question is the infinite momentum frame (IMF) which allows to
interpret the nine {\it space} dimensions in which the D0-branes live as the
nine {\it transverse} dimensions of an eleven dimensional space-time. The
dynamics of $N$ such D0-branes was known to be described by a $N\times N$
matrix quantum mechanics. The BFSS conjecture then is that M-theory in the
IMF is equivalent to  a matrix quantum mechanics of ${\rm U}(N)$ matrices in
the $N\to\infty$ limit, with a particular Hamiltonian that follows from
reducing 9+1 dimensional ${\rm U}(N)$ super Yang-Mills theory to 0+1
dimensions. This conjecture which seems quite bold in the first place passes
several tests. First, BFSS have shown that it contains the Fock space of an
arbitrary number of supergravitons, i.e. massless supergravity multiplets of
256 states, and that it describes the two graviton scattering correctly.
Second, BFSS argue that the matrix model Hamiltonian, always in the 
$N\to\infty$ limit, reduces to the Hamiltonian of the eleven dimensional
supermembrane in the light cone gauge, and hence describes the supermembranes
that must be present in M-theory. Since then, many papers have appeared that
further confirmed this conjecture and elaborated on many other issues in what
has now become known as M(atrix) theory. I will not review these more recent
developments in these lectures.

These lectures are organised as follows: In the first lecture, I introduce and
review D-branes with emphasis on those aspects that will be important to the
M(atrix) theory. Since D0-branes play a particularly important role they
will be given somewhat more attention.  This first lecture is essentially a
selection from Polchinski's excellent TASI lectures \cite{POL}. In particular,
it is shown why a collection of $N$ D$p$-branes is described by a ${\rm U}(N)$
super Yang-Mills theory on the $p+1$ dimensional brane world volume as
obtained by dimensionally reducing ten dimensional super Yang-Mills theory.
For D0-branes this is just quantum mechanics of nine bosonic ${\rm U}(N)$
matrices $X^i$ and their 16 real fermionic partners.
The second lecture then is based on Witten's famous paper \cite{WITM} where it
is shown how eleven dimensional supergravity appears in the strong-coupling
limit of low-energy type IIA superstring theory. Here the Kaluza-Klein modes
of the eleven dimensional supergravity are identified with the D0-branes of
the type IIA superstring.
Then it is practically clear that the Kaluza-Klein modes
of the eleven dimensional supergravity should be described in terms of the
supersymmetric matrix quantum mechanics just mentioned. The third lecture then explains the
conjecture of BFSS that this matrix quantum mechanics actually should
describe the full eleven dimensional M-theory in the infinite momentum frame.

\vskip 8.mm

\section{FIRST LECTURE : D-BRANES}

I will begin by reviewing some basic aspects of D-branes with emphasis one
those that will be important to the matrix model. Most of what will be said in
this lecture can be found in Polchinski's excellent TASI lectures \cite{POL}.

\subsection{\it T-duality for closed strings}

For the closed string the equations of motion $\d_z\d_{\bar z} X^\m=0$ lead
to the expansion 
\begin{eqnarray}
X^\m&=&x^\m  - i\sqrt{\a'\over 2} (\a^\m_0 + \tilde \a^\m_0 ) \t \cr
&{}&\phantom{x^\m }
+ \sqrt{\a'\over 2} (\a^\m_0 - \tilde \a^\m_0 ) \s \cr
&+&i\sqrt{\a'\over 2} \sum_{m\ne 0} 
\left({\a^\m_m\over m}z^{-m}+ {\tilde\a^\m_m\over m}\zb^{-m}\right) 
\label{di}
\end{eqnarray}
where $z={\rm e}^{\t-i\s}$ and $\zb={\rm e}^{\t+i\s}$.
We see that
$p^\m= {1\over \sqrt{2\a'}} (\a^\m_0 + \tilde \a^\m_0 )$. For any
non-compact dimension, invariance of $X^\m$ under $\s\to\s+2\pi$ requires
$\a^\m_0 = \tilde \a^\m_0$.
However, if we compactify one dimension, say $\m=25$ on a circle of radius
$R$, $X^{25} +2\pi R \simeq X^{25}$, and invariance under $\s\to\s+2\pi$  only
requires $\sqrt{\a'\over 2} (\a^\m_0 - \tilde \a^\m_0 ) = m R$ for some
integer $m$. Thus winding states appear. On the other hand, $p^{25}$
must now be quantized as $n/R$. Then
\begin{eqnarray}
\a_0^{25}&=& \sqrt{\a'\over 2} \left( {n\over R} + m {R\over \a'} \right) 
\cr
\tilde\a_0^{25}&=& \sqrt{\a'\over 2} \left( {n\over R} - m {R\over \a'}
\right) \ .
\label{dii}
\end{eqnarray}
The mass of a given string state is given by the sum $(\a_0^{25})^2 +
(\tilde\a_0^{25})^2$ plus contributions of the oscillator modes, and hence is
invariant under flipping the sign of $\tilde\a_0^{25}$ which amounts to
$n\leftrightarrow m$ and $R\leftrightarrow {\a'\over R}$. This is called
T-duality: upon exchanging winding and momentum modes, the theory
compactified on a circle of radius $R$ and the theory compactified on a
circle of radius $\hat R=\a'/R$ are equivalent. It is easy to see that this is
also an invariance of the interacting theory. For this it is enough to 
see that the transformation
\begin{eqnarray}
X^{25} &\equiv& X^{25}(z) + X^{25}(\zb) \cr
\to 
\hat X^{25} &\equiv& X^{25}(z) - X^{25}(\zb) 
\label{diii}
\end{eqnarray}
which changes the signs of all $\tilde\a_m^{25}$ leaves the stress-energy
tensor, all operator product expansions and hence all correlation functions
invariant: T-duality is a symmetry of perturbative closed string theory. It
is a space-time parity operation on the right-moving degrees of freedom only.

\subsection{\it T-duality for open strings}

For an open string, to obtain the equations of motion $\d_z\d_{\bar z} X^\m=0$
upon varying the action, one has to impose either of two types of boundary
conditions. The usual choice are the Neumann (N) conditions 
$\d_{\rm n} X^\m =0$.
But one could just as well impose Dirichlet (D) conditions $X^\m =const$ at
the boundaries. Let's first consider Neumann conditions and later recover the
Dirichlet conditions via T-duality. With the usual N conditions the string
field expansion is
\begin{eqnarray}
X^\m&=&x^\m-i\a' p^\m \ln z\zb \cr
&+&i\sqrt{\a'\over 2} \sum_{m\ne 0} {\a^\m_m\over m}
(z^{-m}+\zb^{-m})  .
\label{div}
\end{eqnarray}
If one compactifies again $X^{25}$ on a circle of radius $R$, one again has
$p^{25}={n\over R}$, but no winding modes can appear. As the radius $R$ is
taken to zero, only the $n=0$ mode survives and the space-time  behaviour of
the open string is as if it lived in one dimension less, although the string
still vibrates in all 26 dimensions (or rather in all transverse 24 ones). It
is similar to what would happen if the endpoints of the open string were
stuck to a hyperplane with $D-1=25$ space-time dimensions.

To understand this
better, one can introduce the T-dual field $\hat X^{25}$ 
by a transformation similar to eq. (\ref{diii}):
\begin{eqnarray}
\hat X^{25}&=&\hat x^{25}-i\a' p^{25} \ln {z\over \zb}\cr
&+&i\sqrt{\a'\over 2}
\sum_{m\ne 0} {\a^{25}_m\over m}(z^{-m}-\zb^{-m}) \ .
\label{dv}
\end{eqnarray}
Instead of the boundary condition $\d_{\rm n}X^{25}\equiv \d_\s X^{25}
=0$ at $\s=0,\pi$ we now have for the T-dual field 
$\d_{\rm t}\hat X^{25}\equiv \d_\t \hat X^{25}=0$ where $\d_{\rm n}$ and
$\d_{\rm t}$ are the normal and tangential derivatives on the boundary of the
string world-sheet which we take to be the infinite strip. The boundary
conditions for $\hat X^{25}$ mean that it is constant along the boundary,
hence these are Dirichlet boundary conditions. The difference between the
values taken by $\hat X^{25}$ at the two boundaries is
\begin{eqnarray}
\hat X^{25}(\pi)-\hat X^{25}(0)&=&2\pi\a' p^{25}=2\pi\a' {n\over R}\cr
&\equiv& 2\pi n \hat R 
\label{dvi}
\end{eqnarray}
where we used the  appropriate T-dual radius $\hat R=\a'/R$. But in the
compactified dual theory $\hat X^{25}$ and $\hat X^{25}+2\pi n \hat R$ 
are to be identified, meaning that both ends of the string lie on  one and
the same $24+1$ dimensional hyperplane.
\begin{figure}[htb]
\vspace{9pt}
\centerline{ \fig{4.5cm}{4.5cm}{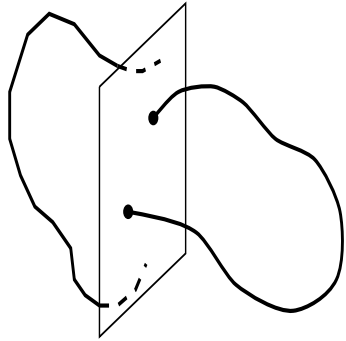} }
\caption{A D$p$-brane is a hypersurface with $p$ space-like and
one time-like dimensions on which open strings with Dirichlet
boundary conditions end.}
\label{singleD}
\end{figure}
The existence of these T-dual open strings is a logical consequence of the
T-duality of the closed string sector {\it contained} in open strings. Thus
these T-dual open strings must be included, i.e. open strings with Dirichlet
boundary conditions on hyperplanes should be included. But now nothing
prevents us from taking $\hat R\to\infty$ showing that these D boundary
conditions should be allowed whether or not $\hat X^{25}$ is compactified or not.
The $24+1$ dimensional hyperplane on which these strings with D boundary
conditions end is called a D-brane, or more precisely a D 24-brane.
T-dualizing more than one dimension, say $k$ space dimensions, leads to D
$p$-branes, with $p=25-k$. A D 25-brane means a $25+1$ dimensional hyperplane:
this just gives back ordinary open strings.

\subsection{\it ${\rm U}(N)$ gauge symmetry}

Open strings can carry Chan-Paton factors at their ends leading to a, say 
${\rm U}(N)$ gauge theory. The open string states then have an additional
label $\vert ij \ra$ with $i,j =1, \ldots N$. Including a background gauge field 
corresponding to  a Wilson  line 
$A_{25}={\rm diag}(\th_1,\th_2, \ldots\th_N)/(2\pi R)$ generically breaks the
gauge symmetry ${\rm U}(N)\to {\rm U}(1)^N$. Now this is pure gauge,
$A_\m=-i U^{-1} \d_\m U$ with 
\begin{equation}
U={\rm diag}\left(\e^{i X^{25} \th_1/2\pi R},
\ldots
\e^{i X^{25} \th_N/2\pi R} \right) 
\label{dvii}
\end{equation}
and can be gauged away by this gauge transformation $U$. However, $U$ is
not periodic under $X^{25}\to X^{25}+2\pi R$. Hence under 
$X^{25}\to X^{25}+2\pi R$ all fields transforming in the fundamental
representation of the gauge group will pick up phases $({\rm e}^{i \th_1},
{\rm e}^{i \th_2},\ldots {\rm e}^{i \th_N})$. Now a state of momentum $p$
picks up a phase ${\rm e}^{ipa}$ under $x\to x+a$, hence we conclude that now
the open string momenta $p^{25}$ may have fractional parts $\sim {\th_i\over
2\pi R}$. More precisely, for an open string whose endpoints are in the state
$\vert ij \ra$ the phase is ${\rm e}^{i (\th_j-\th_i)}$ and possible momenta
are $p^{25}=(2\pi n +\th_j-\th_i)/(2\pi R)$. This corresponds to ``fractional
winding numbers" in the dual picture and, for the positions of the
D-branes on which the T-dual open string can end, leads to
\begin{equation}
\hat X^{25}(\pi)-\hat X^{25}(0)=
(2\pi n +\th_j-\th_i)\hat R \ .
\label{dviii}
\end{equation}
This means that now we do not just have a single D-brane but precisely $N$
D-branes at positions $\th_i \hat R$, see Fig. 2. Note that $\th_i \hat R=2\pi\a' 
{\th_i\over 2\pi R} =2\pi \a' (A_{25})_{ii}$ so that the possible positions are
$(2\pi\a')$ times the eigenvalues of $A_{25}$.
\begin{figure}[htb]
\vspace{9pt}
\centerline{ \fig{4.5cm}{4.5cm}{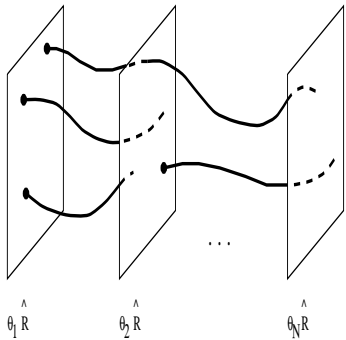} }
\caption{A Wilson line breaking  ${\rm U}(N)$ to ${\rm U}(1)^N$
yields $N$ D-branes at positions  $\th_j \hat R$.}
\label{multipleD}
\end{figure}

In this situation we have $N$ separate D-branes and the gauge group is ${\rm
U}(1)^N$. If we now let  all $\th_i$ coincide the gauge group will no longer be
broken and the full ${\rm U}(N)$ is restored. At the same time, all $N$ D
branes are at coinciding positions, all on top of each other. More generally
we may only take $k$ of the $\th_i$ to be equal, giving ${\rm U}(k)$ for $k$
coinciding D-branes. The lesson is that separated D-branes correspond to a
gauge group which only has ${\rm U}(1)$ factors, while the gauge symmetry is
enhanced to ${\rm U}(k)$  if $k$ branes coincide. This can also be seen by
the following argument. The 25 dimensional mass is
\begin{eqnarray}
M^2&=&(p^{25})^2 +{1\over \a'} ({\cal N}-1) \cr
&=&\left( {(2\pi n +\th_j-\th_i)\hat R\over 2\pi\a'}\right)^2 
+{1\over \a'} ({\cal N}-1) \ .\cr
&{}&
\label{dix}
\end{eqnarray}
Then for the vector states ${\cal N}=1$ with $n=0$, the mass  is given by 
${(\th_j-\th_i)\hat R\over 2\pi\a'}$. This is just the product of the minimal
length of a string streching between the D-brane hyperplanes at
$x^{25}=\th_i\hat R$ and at $x^{25}=\th_j \hat R$ times the string tension
$1/(2\pi\a')$. As $\th_i\to \th_j$  new massless vector states appear,
corresponding to heavy vector bosons becoming massless and thus increasing
the gauge group to ${\rm U}(2)$ times the extra ${\rm U}(1)$ factors.

\subsection{\it Fluctuating D-branes}

Generically (for $\th_i\ne \th_j$) the massless vector states only come from
strings with both ends on the same D-brane. These strings have vertex
operators $V^{(\m)}=\d_{\rm t} X^\m \ ,\ \m=0, \ldots 24$ and
$V^{(25)}=\d_{\rm t} X^{25}=\d_{\rm n} \hat X^{25}$.
The first 25 $V^{(\m)}$ yield gauge fields in the D-brane while 
$V^{(25)}$, due to the appearance of the normal derivative describes
transverse fluctuations of the brane. How can this be since we started with
a rigid hyperplane? The mechanism is familiar in string theory where one
starts e.g. in Minkowski space and then discovers that for the closed
string the massless modes describe fluctuations of the  space-time geometry.
Here $\d_{\rm n} \hat X^{25}$ similarly describes the transverse fluctuations
of the brane. As a result the D-brane becomes dynamical.

\subsection{\it D-brane actions}

What is the effective action induced on the D$p$-brane  world-volume that
effectively describes low-energy processes of D-branes? One proceeds in
exactly the same way as for determining the closed string effective action,
except that now boundary terms must be taken into account. We start with a
standard $\s$-model action in the bulk of the open string world sheet and
then add the appropriate boundary couplings ($\delta_i$ are the displacement fields
corresponding to the dualized components of $A_\m$):
\begin{eqnarray}
S_{\rm boundary}&\sim &\int {\rm d}s \sum_{m=0}^p A_m(x^0, \ldots x^p) \d_{\rm
t}X^m \cr
&+&\int {\rm d}s \sum_{i=p+1}^{25} 
\delta_i (x^0, \ldots x^p) \d_{\rm n} \hat X^i .\cr
&{}&
\label{dx}
\end{eqnarray}
Note that the fields $A_m,\ \delta_i$ only depend on the zero-modes $x^0,
\ldots x^p$ that are in the brane.
Requiring the sum of bulk and boundary $\s$-model to be conformally invariant
leads to the $\b$-function equations for $\f$, $G_{\m\n}$, $B_{\m\n}$ as well
as for $A_m$ and $\delta_i$. Then as usual, these equations can be obtained
by varying a certain action functional that is interpreted as the effective
action we are looking for. It is given by
\begin{eqnarray}
S_{{\rm D}p{\rm brane}}^{\rm effective}
&=&-T_p \int {\rm d}^{p+1}\xi\,
\e^{-\f}\times\cr
&\times&\left[\det\left( g_{mn} +b_{mn} 
+2\pi\a'F_{mn}\right)\right]^{1/2}\cr
&{}&
\label{dxi}
\end{eqnarray}
where $g_{mn}=(\d X^\m/\d\xi^m)(\d X^\n/\d\xi^n) G_{\m\n}$, etc are the
pull-backs of the space-time fields 
$G_{\m\n},\ B_{\m\n}$ to the brane, and $T_p$ is the D$p$-brane tension given
below. For trivial metric and antisymmetric
tensor background as well as constant dilaton field ($\e^\f=g_s$) 
this is just the Born-Infeld action computed long ago in
ref. \cite{ACNY}. 
If moreover one expands this action to lowest non-trivial order in $F$ one gets up
to a constant
\begin{equation}
S_{{\rm D}p}^{\rm eff} \to - {T_p\over g_s}   \int {\rm d}^{p+1}\xi\,
{1\over 4} (2\pi\a')^2 F_{mn}^2 \ .
\label{dxia}
\end{equation}
The action (\ref{dxi}) or (\ref{dxia}) is valid for a {\it single}
D$p$-brane. If there are $N$ such branes, extra terms appear as we will see
next.

\vskip 5.mm

\subsection{\it D-brane coordinates as non-commuting matrices}

That D-brane coordinates should be represented by matrices might seem strange
at first sight. But, following Witten \cite{WITP}, I will now show that this
follows very naturally from the properties of D-branes I already explained.
We have seen that $N$ D-branes correspond to a gauge group ${\rm U}(1)^N
\subset {\rm U}(N)$ with the massless vector states being $a_{-1}^\m \otimes
\vert ii \ra$. As these D-branes coincide, one recovers the full gauge group
${\rm U}(N)$. As we have seen, for separated D-branes the breaking of ${\rm
U}(N)$ is due to the mass terms associated with the off-diagonal
$(A_\m)_{ij}$ components
which are given by the product of the string tension and the distance between
the branes $i$ and $j$. For $N$ coinciding branes, the low-energy effective
action must be the non-abelian ${\rm U}(N)$ Yang-Mills theory reduced to the
brane world-volume. But the world-volume is the same for all $N$ branes,
coinciding or not, so the non-abelian ${\rm U}(N)$ YM theory must remain the
correct effective action even for separated D-branes.

The point now is that
while $A_m$ with $m=0,1, \ldots p$ are actually the gauge fields that live on
the brane, the remaining components $A_i$ with $i=p+1, \ldots 25$ describe the
transverse fluctuations, i.e. the positions of the branes. Before, for a
single D-brane, we called them $\delta_i$ but now it seems more appropriate
to call them $X_i$. 
More precisely, after eq. (\ref{dviii}) we have seen that the correct
normalisation includes a factor $2\pi\a'$: 
\begin{equation}
A_i={1\over 2\pi\a'} X_i \ .
\label{dxib}
\end{equation}
Since the $X_i$ are just the components of the 26
dimensional gauge field normal to the brane, it is clear that they, too, must
be ${\rm U}(N)$ matrices. We will see shortly that for widely separated D
branes the eigenvalues of $X_i$ are just the coordinates of the $N$ D-branes
while the off-diagonal elements take into account the interactions that arise
upon  integrating out the open strings connecting two different D-branes.

The effective action is just ten dimensional super Yang-Mills theory reduced
to $p+1$ dimensions:
\begin{eqnarray}
S_{\rm YM}^{(p+1)}&\sim& -{T_p (2\pi\a')^2\over 4  g_s} \times \cr
&\times &\int {\rm d}^{p+1}\xi\, \tr \left( F_{mn}^2 +2
F_{mj}^2 +F_{ij}^2\right) 
\label{dxii}
\end{eqnarray}
where the overall normalisation is consistent with (\ref{dxi}) and (\ref{dxia}).
I use a metric of
signature $(-+\ldots +)$ and e.g. $F_{mj}^2$ is meant to be 
$\sum_{j=p+1}^{25} \left( -F_{0j}^2 +\sum_{n=1}^p F_{nj}^2 \right)$, etc.
On the D-brane there is no dependence on the zero-modes 
$x^i$, $i=p+1, \ldots 25$ because
the D boundary condition has removed the zero-modes in the directions normal
to the brane. This means that all derivatives in the $i$ directions disappear.
Hence 
\begin{eqnarray}
F_{mn}&=&{\d A_n\over \d x^m}-{\d A_m\over \d x^n}+i[A_m,A_n]\cr
&{}& \cr
(2\pi\a')F_{mj}&=&{\d X_j\over \d x^m}+i[A_m,X_j]\equiv D_m X_j\cr
&{}& \cr
(2\pi\a')^2 F_{ij}&=& i[X_i,X_j] \ .
\label{dxiii}
\end{eqnarray}
Upon inserting this into the action (\ref{dxii}) we get 
\begin{eqnarray}
&{}&S_{\rm YM}^{(p+1)}\sim -{T_p (2\pi\a')^2\over 4 g_s}\int {\rm d}^{p+1}\xi\, 
\tr F_{mn}^2\cr
&{}&+ {T_p\over g_s}
\int {\rm d}^{p+1}\xi\, \tr \Big( -{1\over 2} (D_m X^i)^2 \cr
&{}& \phantom{ {T_p\over g_s}\int {\rm d}^{p+1}\xi\, \tr \Big(   } 
+{1\over 4}
{([X^i,X^j])^2\over (2\pi\a')^2}\Big) \ .
\label{dxiv}
\end{eqnarray}
The first term is just the $p+1$ dimensional YM action on the brane and is
the low-energy limit of the obvious non-abelian generalisation of the
Born-Infeld effective action we discussed above for a single brane. The
second and third term are the effective action governing the D-brane
dynamics. In a superstring theory there will be additional fermionic terms.
In any case, the scalars $X_i$ have a potential\footnote{
Note that this is a non-negative potential: since $X^i$ and $X^j$ are
hermitian, $i [X^i,X^j]$ is also hermitian, so that the potential is the square
of a hermitian matrix, hence non-negative.
}  
energy $\sim -\tr ([X^i,X^j])^2$.
In the supersymmetric case,  a vacuum with unbroken
supersymmetry must have vanishing potential and thus $[X^i,X^j]=0\ \forall \
i,j$. Then all matrices $X^i$ are simultaneously diagonalisable:
$X^i={\rm diag} (a^i_{(1)}, \ldots a^i_{(N)})$. We interpret these
eigenvalues $a_{(k)}^i$ ($i=p+1, \ldots 25$) as giving the coordinates
of the $k^{\rm th}$ brane. Expanding around these vacua gives masses for the
off-diagonal modes $\sim {1\over 2\pi\a'}\vert a_{(k)}-a_{(l)}\vert$ as we
expect from the string mass formula that yielded masses proportional to the
distance between brane $k$ and brane $l$. The collection of the $a_{(k)}^i$ 
parametrizes the moduli space of ${\rm U}(N)$ susy vacua rather than really
giving the positions of the branes in the quantum theory. When the branes are
nearby, many mass terms are small and the massive modes cannot be neglected
and one must study the full ${\rm U}(N)$ YM theory. This is the origin of the
non-commuting matrix character of the ``positions" $X^i$ of the D-branes.

\vskip 5.mm

\subsection{\it D-branes in superstrings}

In type IIA or type IIB superstring theories it is similarly natural to
introduce D-branes on which open type I superstrings can end. These D-branes
couple naturally to the $p+1$ form RR gauge field. Indeed, D-branes are
invariant under half the supersymmetries and hence are BPS states. Thus they
must carry conserved abelian charges which are the RR charges in question.
The IIA theory has RR gauge fields $A_\m,\ A_{\m\n\r}, \ A_{\m\n\r\s\lambda}$
so they couple to D0-branes, D2-branes, D4-branes, etc.
The IIB theory has RR gauge fields $A,\ A_{\m\n}, \ A_{\m\n\r\s}$
so they couple to D(-1)-branes, D1-branes, D3-branes, etc. 
The D(-1)-branes are D instantons, having also a D boundary condition in the time
direction, D0-branes are D particles, while the D1-brane is also called a D
string and the D2-brane a D membrane.

Everything we have seen before remains valid with the obvious
supersymmetric modifications. In particular, the effective YM action is now
replaced by a super YM action which also includes the 16 real component
spinors $\p$, also in the adjoint representation of ${\rm U}(N)$. There is
now also a term describing explicitly the couplings of the 
D-branes to the RR gauge fields.
One can compute the various coefficients in front of the effective D-brane
actions: the D$p$-brane tension $T_p$ and charge $\m_p$. They are given by
the appropriate disc diagrams.

I will need the explicit expressions for the tensions. Following the notation of
\cite{POL}, $T_p$ denotes the tension 
without the factor of $1/g_s$ while it is included in
$\tau_p$:
\begin{equation}
T_p={ (2\pi\sqrt{\a'})^{1-p} \over 2\pi\a'} \ ,\ \tau_p={T_p\over g_s} \ .
\label{dxiva}
\end{equation}
In particular one has
\begin{equation}
T_0={1 \over\sqrt{\a'}}\equiv{1\over l_s} \ .
\label{dxivb}
\end{equation}

\subsection{\it Effective D0-brane action}

We have seen that the effective D$p$-brane action is ten dimensional super
YM theory dimensionally reduced to $p+1$ dimensions. For $p=0$ this gives
\begin{eqnarray}
S^{\rm D0}&=& T_0 \int {\rm d}t\, 
\tr \Big( -{1\over 4 g_s c^2} F_{\m\n}F^{\m\n}\cr
&{}& \phantom{T_0 \int {\rm d}t\,  \tr \Big( }
+i\bar\Psi \Gamma^\m D_\m \Psi\Big) 
\label{dxv}
\end{eqnarray}
where here and in the following I write
\begin{equation}
c={1\over 2\pi \a'}
\label{dxva}
\end{equation}
for short.
The factor of $1/g_s$ in the effective action comes from computing a disc
diagram. A corresponding factor for the fermions has been reabsorbed in the
normalisation of $\Psi$. The general representation of the Clifford algebra
is 32 dimensional, but $\Psi$ is a Majorana-Weyl spinor $\Psi=\pmatrix{
\th\cr 0\cr}$ with $\th$ being a real 16 component spinor. 
We take the $\Gamma^\m$ to be
\begin{equation}
\Gamma^0=\pmatrix{0&-1\cr 1&0\cr} \ , \ 
\Gamma^j=\pmatrix{0&\g^j\cr \g^j&0\cr}
\label{dxvi}
\end{equation}
where $\g^i$, $i=1,\ldots 9$ are real symmetric $16\times 16$ gamma matrices
of ${\rm SO}(9)$. We have again $F_{ij}=i c^2 [X^i,X^j]$, $F_{0j}=c D_0 X^j\equiv
c \d_0 X^j +i c [A_0,X^j]$ as well as $D_j\th=i c [X_j,\th]$ and
$D_0\th=\d_0\th+i[A_0,\th]$
so that
\begin{eqnarray}
S^{\rm D0}&=& T_0 \int {\rm d}t\, \tr \Big( 
{1\over 2 g_s} (D_0X^i)^2 - i\th^T D_0 \th\cr
&+&{c^2\over 4 g_s}([X^i,X^j])^2 + c \th^T \g^j [X_j,\th] \Big) \ .
\label{dxvii}
\end{eqnarray}
This is a supersymmetric quantum mechanics for $X^i$ and $\th$ in the adjoint
of ${\rm U}(N)$, i.e. $N\times N$ hermitian matrices, and each component
of the $\th$ matrix is a real 16 component spinor.

\section{SECOND LECTURE : THE APPEARANCE OF THE ELEVENTH DIMENSION -
M-THEORY}

The low-energy effective theory of type IIA super{\it string} is ten
dimensional type IIA super{\it gravity}. Type IIA super{\it gravity} can also be
obtained by dimensional reduction of super{\it gravity} in eleven dimensions.
Since long, this had prompted the question of whether eleven dimensional
supergravity is some low-energy effective theory of some consistent quantum
theory in eleven dimensions. For some time it had been hoped that this might
be a theory with supermembranes as its fundamental objects. 

In this lecture,
following Witten  \cite{WITM}, I will argue that there is indeed such an eleven
dimensional theory, called M-theory which could be viewed as the
strong-coupling limit of the ten dimensional type IIA superstring.

\subsection{\it Supergravity in eleven and IIA supergravity in ten dimensions}

The eleven dimensional supergravity multiplet contains the following massless
fields: a metric $G_{MN}$ or equivalently an elevenbein $e_M^A$, a three-form
potential $A_3$ with components $A_{MNP}$ and a Majorana gravitino $\P_M$. To
count the physical degrees of freedom of massless fields in dimension $d$, the
simple rule is to do the counting as if one were in $d-2$ dimensions and all
components were physical. Hence a symmetric traceless tensor like the metric
has  ${1\over 2}(d-2)(d-1)-1$ physical degrees of freedom (dofs). 
For $d=4$ this gives
the familiar two dofs, while it gives 35 dofs in ten dimensions 
and 44 dofs for the
eleven dimensional $G_{MN}$. The antisymmetric three-index tensor has ${1\over
6}(d-2)(d-3)(d-4)$ dofs, which gives 84 dofs for $d=11$. This makes a total of
128 bosonic dofs for eleven dimensional supergravity. For the fermionic
partners the counting is similar: the eleven dimensional Clifford algebra has
32 dimensional spinors. Imposing a Majorana condition gives 32 component
real spinors. Due to the Dirac equation only half of them are physical, so a
Majorana spinor has 16 real dofs. The gravitino also has a vector index, which
contributes a facor of $d-2=9$. However one has to project out the spin
${1\over 2}$ components, which leaves us with $16\times 9-16=128$ (real)
fermionic dofs. The bosonic part of the eleven dimensional supergravity action
is schematically (in units where $\a'=1$)
\begin{eqnarray}
S_{\rm bos}^{(11)}&=& {1\over 2}\int {\rm d}^{11}x\, \sqrt{G}\left( R +\vert
{\rm d} A_3 \vert^2\right)\cr
&+& \int A_3\wedge {\rm d} A_3 \wedge {\rm d} A_3 
\label{ti}
\end{eqnarray}
with the fermionic terms determined by supersymmetry. Here and in the rest of
this subsection, I do not care about the precise numerical factors in front of
each term in the action.

Next, one reduces this theory to ten dimensions (indices $\m,\n,\r$), i.e. one
takes $x^{11}$ on a circle and supposes that nothing depends on $x^{11}$. 
The eleven dimensional Majorana gravitino 
$\P_M\equiv \pmatrix{ \p^1_M\cr \p^2_M}$ gives rise in ten dimensions to a
pair of Majorana-Weyl gravitinos (of opposite chirality) 
$\p^a_\m$ and a pair of
Majorana-Weyl spinors $\p^a\equiv \p^a_{11}$, $a=1,2$.
The
eleven dimensional three-form gives rise in ten dimensions to  a three form
$A_{\m\n\r}$ (56 dofs) and a two-form $B_{\m\n}\equiv A_{\m\n 11}$ (28 dofs),
while the eleven dimensional metric gives in ten dimensions a metric\footnote{
To be precise, as one can see from the next equation, $G_{\m\n}$ is not $G_{MN}$ for
$M=\m$ and $N=\n$, but it also contains an additional term $-\e^{2\g}A_\m A_\n$.
} 
$G_{\m\n}$
(35 dofs),  a scalar $\e^{2\g}\equiv G_{11\, 11}$, and a vector
potential (one form) $A_\m\equiv - \e^{-2\g} G_{\m\, 11}$ (8 dofs), again a
total of 128 bosonic dofs. An important point concerns the interpretation of
$\e^{2\g}$. We take $x^{11}$ to vary from 0 to $2\pi$. However this does not
fix the size of the compact dimension  because the eleven dimensional line
element is 
\begin{eqnarray}
{\rm d}s^2&=& G_{MN} {\rm d}x^M {\rm d}x^N \cr
&=& G_{\m\n} {\rm d}x^\m {\rm d}x^\n + \e^{2\g} 
({\rm d}x^{11}-A_\m {\rm d}x^\m)^2
\label{tii}
\end{eqnarray}
and one sees that the eleventh dimension is a circle of radius $\e^\g$.
Equation (\ref{tii}) also shows that $\det G_{(11)}=\e^{2\g} \det G_{(10)}$ and
thus
\begin{equation}
\int {\rm d}^{11}x \sqrt{G_{(11)}}\ldots = 2\pi \int 
{\rm d}^{10}x\, \e^\g\sqrt{G_{(10)}}\ldots
\label{tiii}
\end{equation}
For the bosonic action (\ref{ti})
one gets (remember that I do not care about numerical
factors)
\begin{eqnarray}
&{}& \int {\rm d}^{10}x\, \sqrt{G_{(10)}} \Big[ \e^\g 
(R +\vert \N\g\vert^2 +\vert {\rm d} A_3\vert^2)\cr
&{}& \phantom{\int {\rm d}^{10}x\, \sqrt{G_{(10)}} \Big[}
+\e^{3\g} \vert {\rm d} A\vert^2 + \e^{-\g} \vert {\rm d} B\vert^2 \Big]\cr
&{}& +\int B\wedge {\rm d}A_3 \wedge {\rm d}A_3 \ .
\label{tiv}
\end{eqnarray}
Let me say a word about where the powers of $\e^\g$ come from. We have already
seen that the square-root of the determinant of the metric gives a factor of
$\e^\g$. The eleven dimensional curvature gives rise to terms $G^{11\, 11}
\d_\m G_{\n\, 11}\d_\r G_{\s\, 11} \sim \e^{-2\g} \e^{2\g} \e^{2\g} \d_\m
A_\n \d_\r A_\s$ plus $(\d\g)$-terms. This gives $\
e^{2\g}\d A\d A$, and together with the $\e^\g$ from the determinant a factor
of $\e^{3\g}$ in front of $\vert {\rm d} A\vert^2$. Similarly, to get 
$\vert {\rm d} B\vert^2$ from $\vert {\rm d} A_3\vert^2$ one needs to consider
$G^{11\, 11} \d A_{\m\n\, 11}\d A_{\r\s\, 11} 
\sim \e^{-2\g} \vert {\rm d} B\vert^2$. Together with the $\e^\g$ from the
determinant this gives a factor of $\e^{-\g}$.

The action (\ref{tiv}) contains all terms one wants to obtain for IIA
supergravity in ten dimensions. The factors of $\e^\g$, however, 
are not what one expects. The usual form of the action for IIA supergravity is
(again not worrying about numerical factors)
\begin{eqnarray}
&{}&\int {\rm d}^{10}x\, \sqrt{g} \Big[ \e^{-2\f} 
(R +\vert \N\g\vert^2 +\vert {\rm d} B\vert^2)\cr
&{}& \phantom{\int {\rm d}^{10}x\, \sqrt{g} \Big[}
+ \vert {\rm d} A_3\vert^2 + \vert {\rm d} A\vert^2 \Big]\cr
&{}& +\int B\wedge {\rm d}A_3 \wedge {\rm d}A_3 \ .
\label{tv}
\end{eqnarray}
To bring the action (\ref{tiv}) in the form (\ref{tv}) all one needs to do is to
perform a Weyl rescaling of the metric:
\begin{equation}
G_{\m\n}=\e^{-\g} g_{\m\n}\ .
\label{tvi}
\end{equation}
It then follows that $\sqrt{G_{(10)}}=\e^{-5\g} \sqrt{g}$, while
$R[G_{(10)}]=\e^\g R[g]$ plus terms $\sim \vert\N\f\vert^2$. For a $p$ form
$A_p$ the kinetic term $\vert {\rm d} A_p\vert^2$ contains $p+1$ inverse metrics
$G^{\m\n}$ and hence gives  $\e^{(p+1)\g}$ times
$\vert {\rm d} A_p\vert^2$ with indices now contracted using $g^{\m\n}$. Hence
we see that the action (\ref{tiv}) takes the desired form if we identify
$\e^{-3\g}=\e^{-2\f}$, i.e. 
\begin{equation}
\e^{\g}=\e^{2\f/3} \ .
\label{tvii}
\end{equation}

\subsection{\it String coupling, radius and KK modes}

Recall that $\e^\g$ was the radius of the eleventh dimension, or putting back
$\a'$ one has $R_{11}=\sqrt{\a'} \e^\g$. On the other hand, $\f$ being the
dilaton, $\e^\f$ is the  coupling constant. In string theory it is the string
coupling constant $g_s$, so that one arrives at the relation 
\begin{equation}
R_{11}=\sqrt{\a'} g_s^{2/3} \ .
\label{tviii}
\end{equation}
One has to be a bit careful. This radius is the radius of the eleventh
dimension when measured with the eleven dimensional metric $G$. If we measure
distances instead with the Weyl rescaled (string) 
metric $g$ one has ${\rm d}s^2_{g} =
\e^\g {\rm d}s^2_{G} = \left(\e^{\f/3}\right)^2 {\rm d}s^2_{G}$ and thus\footnote{
Do not confuse $g$, which designs the string metric, with $g_s$ which is the string
coupling constant!}
\begin{equation}
R_{11}^{(g)}=\e^{\f/3} R_{11}^{(G)} = g_s^{1/3} R_{11}^{(G)} =\sqrt{\a'} g_s \ .
\label{tix}
\end{equation}
Introducing the string length $l_s=\sqrt{\a'}$ and string mass
$m_s=1/\sqrt{\a'}=1/l_s$, this can also be written as
\begin{equation}
R_{11}^{(g)}=g_s l_s \ .
\label{tx}
\end{equation}

When the eleven dimensional supergravity is compactified on a circle of radius
$R_{11}$, the resulting ten dimensional theory not only has the massless modes
described so far that form a supergravity multiplet with 256 states (128
bosonic and 128 fermionic) but there also are all the massive Kaluza-Klein (KK) modes.
Since the compactification radius is $R_{11}$ these KK modes have momenta
$n/R_{11}$. When the momenta are measured with the metric $g$ they are of
course $n/R_{11}^{(g)}$. Since the eleven dimensional states are massless, 
this KK momentum is the only contribution to
the ten dimensional mass which thus simply is
\begin{equation}
M={n\over R_{11}^{(g)}} = {n\over \sqrt{\a'} g_s} = {n\over l_s g_s} \ .
\label{txi}
\end{equation}
Each of these massive KK states is a supermultiplet of 256 states. The
important point is that for fixed $n$ their masses are proportional to the
inverse of the string coupling constant, so that in the limit of large string
coupling they become very light.

\subsection{\it The strong-coupling limit of the IIA superstring}

Let us now consider the low-energy sector of type IIA superstring theory. Its
effective action is given  by eq. (\ref{tv}). The field $\f$ is the
dilaton and $\e^\f$ or rather its expectation value is the string coupling
constant $g_s$. In particular we see that the action is such that all
kinetic terms for the NS-NS fields, i.e. the metric, dilaton and two-form
potential, have a factor of $g_s^{-2}=\e^{-2\f}$ in front. The kinetic terms
for the RR fields, i.e the one-form and three-form potentials $A$ and $A_3$,
have no $\f$ dependence. This ``normalisation" of the RR fields is fixed by
requiring that their gauge transformation laws be $\f$-independent.
Concentrate now on the one-form\footnote{
One might ask why not consider the three-form instead. 
This is simply because $A_3$
couples to D2-branes and their masses being equal to their tension times
their area, they do not have the same ``universal" mass as the D0-branes and
thus seem not suited for the identification with the KK states one has in mind.
} 
potential $A$. There are no states in the {\it
perturbative} string spectrum that carry a charge for $A$. These charges are
carried by the D0-branes, as we saw that a D$p$-brane naturally couples to a
$p+1$ form potential via $\int_{p-{\rm brane}} A_{p+1}$. As noted in the
previous section, the corresponding charge $\m_p$ can be computed by a string 
diagram and is proportional to $T_p$. 
We also saw that D$p$-branes are BPS states. 
BPS states break half of the supersymmetries and
they saturate the bound on the masses, i.e. $M=\vert Z\vert$ where $Z$ is the
central charge of the ${\cal N}=2$ supersymmetry algebra. This algebra in ten
dimension schematically reads $\{Q,Q\}\sim \{Q',Q'\}\sim P$ and $\{Q,Q'\}
\sim Z$. If this algebra is derived from eleven dimension, the central charge
$Z$ just is the eleventh component of the momentum $P$. Now the central
charge must be made up from the abelian charges, here the RR charge of the D0-brane. Indeed, it is not too difficult to work out the susy algebra for D
$p$-branes. For a single D0-brane one finds $Z={T_0\over g_s}=\t_0$ 
where $\t_p$ denotes
the D$p$-brane tensions, cf eqs. (\ref{dxiva},\ref{dxivb}). For the D0-brane  this
gives $Z=\t_0={1\over \sqrt{\a'} g_s}$ and hence
\begin{equation}
M={1\over \sqrt{\a'} g_s} = {1\over l_s g_s} \ .
\label{txii}
\end{equation}
So the type IIA superstring contains states, the D0-branes of masses
$M=1/(l_s g_s)$ that come in short supermultiplets of $2^8=256$ states.
It has been shown \cite{WITP,SETHI,PORR}
that a system of $n$ D0-branes has a bound state at
threshold, i.e. this bound state has a total mass of exactly ${n\over l_s g_s}$.
Again this is  a full supermultiplet of 256 states. In the strong
coupling limit $g_s\to\infty$ all these states become very light, and we get
infinitely many light states. This  is {\it exactly} the Kaluza-Klein
spectrum of eleven dimensional supergravity we discussed above. This is quite
surprising!
One knew already that simple dimensional reduction of eleven dimensional
supergravity gives the massless IIA supergravity in ten dimensions. What
Witten has shown, and what I explained here, is that {\it all} KK states of
the full eleven dimensional supergravity on ${\bf R}^{10}\times S^1$ are
contained in the IIA superstring, with each ``state" being actually a full
supermultiplet of 256 states. Moreover, in the strong-coupling limit
$g_s\to\infty$ these are {\it all} low-energy states of the IIA superstring.
Also, for $g_s\to\infty$ one has $R_{11}\to\infty$ and one gets
uncompactified eleven dimensional supergravity:
Eleven dimensional supergravity is the low-energy limit of IIA
superstring theory at strong coupling with $R_{11}^{(g)}\sim g_s$.
An eleventh dimension has been constructed out of the ten dimensional theory
in an entirely non-perturbative way. Clearly, it will be quite non-trivial to
see how this eleven dimensional theory, in the uncompactified limit, can
manage to have eleven dimensional Lorentz invariance.

Of course, eleven dimensional supergravity is not expected to yield a
consistent quantum theory. It should only be the low-energy limit of some
consistent theory, baptised M-theory. The latter should then also  describe
the strong-coupling limit of IIA superstrings, not only at low energies. More
precisely, M-theory with its eleventh dimension compactified on a circle of
radius $R_{11}$ should be identical to IIA superstring theory with string
coupling $g_s=R_{11}^{(g)}/\sqrt{\a'}$ where $R_{11}^{(g)} = g_s^{1/3}
R_{11}$ is the eleven dimensional radius when measured with the string
metric $g$. This can be taken as the definition of M-theory. What else do
we know about it?
Since it should describe IIA superstrings which have D0, D2, D4, D6 and D8
branes as well as the solitonic 5-brane and the fundamental string (F1
brane), M-theory should also contain extended objects. Since the higher
dimensional D$p$-branes are in a certain way dual to the lower dimensional
ones, one mainly has to worry about the latter. If M-theory contains
pointlike degrees of freedom, as well as membranes (i.e. 2-branes) and
5-branes, then things work out. Indeed, the extended $p$-branes of M-theory
may or may not be wrapped around the compact $S^1$, hence yielding $p-1$ or
$p$ branes in the ten dimensional superstring theory. This  gives 
the branes of IIA superstrings with $p=0,1,2,4,5$ as it should. The D6 and
D8 branes are more subtle.

\section{THIRD LECTURE : A MATRIX-MODEL FOR M-THEORY IN THE IMF}

This section is based mainly on the BFSS paper \cite{BFSS}.
So far we have seen that 
a) 
in the strong-coupling limit of IIA superstrings an eleventh dimension
appears and that the KK states of eleven dimensional supergravity
correspond to bound states at threshold of $n$
D0-branes, and
b)
a collection of $n$ D0-brane is described by ten dimensional ${\rm
U}(n)$ super Yang-Mills theory reduced to
$0+1$ dimensions, i.e. by $n\times n$ hermitian matrix quantum mechanics.
\par

There still seems to be a mismatch between the ten dimensions of the matrix
model and the eleven dimensions of the supergravity. 
Here comes the third idea: 
c)
the main idea of BFSS is to interpret the 9 {\it space}
dimensions (the $X^i,\ i=1, \ldots 9$) of the D0-brane matrix model
as the {\it transverse} dimensions of an eleven dimensional theory in the
light-cone frame, or more precisely in the infinite momentum frame (IMF). We
are familiar with the light-cone quantization of the ordinary string
where the light-cone Hamiltonian is $H={1\over p^+} {\cal H}_\perp$. There,
${\cal H}_\perp$ is a Hamiltonian for the $d-2$ transverse degrees of freedom.
The string theory is nevertheless Lorentz invariant in all $d$ dimensions
(provided $d=d_{\rm crit}$). If one manages to interpret the matrix quantum
mechanical Hamiltonian for the nine $X^i$ as a transverse Hamiltonian, then
the full system lives in 11 dimensions and should exhibit eleven
dimensional Lorentz invariance. This is the way how the mismatch of
dimensions can be resolved. 
It is thus useful to first recall some facts about the
infinite momentum frame.

\subsection{\it The infinite momentum frame (IMF)}

The infinite momentum frame (IMF) was introduced in quantum field theory by
Weinberg long ago \cite{WEIN} as a mean to simplify perturbation theory.
Perturbation theory in the IMF is characterised by the vacuum being
trivial. In particular, Feynman diagrams with
vertices where   particles are created out of the vacuum are vanishing in
the IMF. In this sense, the IMF perturbation theory of QFT looks much like
the ``old-fashioned" non-covariant perturbation theory, but with the
energy denominators replaced by covariant denominators.

For a collection of particles, the IMF is defined to be a reference frame
in which the total momentum $P$ is very large. All individual momenta can be
written as
\begin{equation}
p_a=\eta_a P + p_\perp^a
\label{qi}
\end{equation}
with $p_\perp^a\cdot P=0$, $\sum p_\perp^a=0$ and $\sum \eta_a=1$. This means that the
observer is moving with high velocity in the $-P$ direction. If the system
is boosted sufficiently, all $\eta_a$ are positive. For massive particicles
it is obvious that a sufficiently large boost in the $P$ direction will
eventually make all components of the momenta in the longitudinal $P$
direction positive. For a massless particle the same will be true except
if it moves exactly in the $-P$ direction with all transverse $q_a$
vanishing. But this latter case is somewhat degenerate and can simply be
avoided by not boosting the system in exactly the opposite of the direction
of the momentum of any massless particle.\footnote{
Of course, for a {\it single} massless particle, boosting in a direction
$P$ that does not coincide with its momentum does not allow to impose the
condition $p_\perp\cdot P=0$, but this is not crucial. For a system of
particles (other than all massless and with exactly aligned momenta) one
can always first go to the center of mass frame where $p_\perp^a\cdot P=0$
and $\sum p_\perp^a =0$ 
and then boost in any
desired direction to achieve $\eta_a>0$. Obviously then
$p_\perp^a\cdot P=0$ and $\sum p_\perp^a =0$
remain valid.
} 
Hence we can assume that with a
sufficiently large boost {\it all} $\eta_a$ are strictly positive. 
Once $P$ is large enough, further boosting only
increases the total momentum $P$ but does not change the $\eta_a$ anymore,
and of course, the $q_a$ aren't changed either. The
energy of any particle is
\begin{eqnarray}
E_a&=&\sqrt{p_a^2+m_a^2}\cr
&=&\eta_a P 
+{(p_\perp^a)^2 +m_a^2\over 2 \eta_a P} +{\cal
O}(P^{-2}) \ .
\label{qii}
\end{eqnarray}
Apart from the constant $\eta_a P+{m_a^2\over 2 \eta_a P}$ this has the
non-relativistic structure ${(p_\perp^a)^2\over 2 \m_a}$ of a $d-2$
dimensional system with the role of the
non-relativistic masses $\m_a$ played by $\eta_a P$. 

Let us now turn to quantum
field theory. Then internal lines in Feynman diagrams can carry arbitrary
large momenta and for part of the integration range one does not have
$\eta_a>0$. Weinberg has shown, starting from ``old-fashioned" perturbation
theory with energy denominators that whenever an internal $\eta_i$ is
negative the corresponding diagram is suppressed by extra factors of $1/P$.
These suppressed diagrams correspond exactly to diagrams with vertices
where several particles are created from the vacuum. It is in this sense
that in the IMF the vacuum has no non-trivial structure. Hence we conclude
that in field theory also the internal lines have momenta with $\eta_a>0$
only.

It might be useful to compare the IMF to a standard light cone frame. In the
latter one again singles out one spatial direction called longitudinal with
momentum $p_L^a=\eta_a P$ and defines 
$p_{\pm}^a = E^a\pm p_L^a = E^a\pm \eta^a P$. Then the mass shell condition
reads $p_-^a p_+^a - (p_\perp^a)^2=m_a^2$ or
\begin{equation}
E_a -\eta_a P= {(p_\perp^a)^2 +m_a^2\over p_+^a}
\label{qiia}
\end{equation}
This is exact whether $p_L^a$ is large or not. However, if $P$ and hence
$p_L^a=\eta_a P$ is large, one has $E^a \simeq \eta_a P$ and $p_+^a\simeq 2
\eta_a P$ so that one recovers eq. (\ref{qii}).

\subsection{\it M-theory in the IMF}

Now we will consider M-theory in the IMF. We will separate the components
of the eleven dimensional momenta as follows: $p_0,\ p_i,\ i=1,\ldots 9$ and
$\pe$. The $p_i$ will collectively be called $p_\perp$.  
We boost in the $11$ direction to the IMF until all momenta in
this direction are much larger than any relevant scale in the problem. In
particular all $\pe^a$ are stricly positive. We also compactify $x^{11}$
on a circle of radius $R$ (we no longer write the subscript ``11" because
throughout these notes no other dimension will be compactified). To be
precise, when I write $R$, I mean $R_{11}^{(g)}$. All
momenta $\pe^a$ are now quantized as $n_a/R$ with $n_a>0$. Since there are
no eleven dimensional masses $m_a$ the energy dispersion relation just is
\begin{equation}
E-\pe^{\rm tot} = \sum_a {(p_\perp^a)^2\over 2 \pe^a} \ .
\label{qiii}
\end{equation}
Again we see the non-relativistic structure. It exhibits full Galilean
invariance in the transverse dimensions. Actually, M-theory in the IMF
must have
super-Galilean invariance because there are also the 32 real
supersymmetry generators. In the IMF, they split into two groups of 16, 
each transforming as a spinor of ${\rm SO}(9)$: the $Q_\a$ and $q_A$, $\a,
A=1, \ldots 16$ satisfy the algebra
\begin{eqnarray}
\{ Q_\a,Q_\b\}=\delta_{\a\b} H &,&
\{ q_A,q_B\}=\delta_{AB} P_{11}\cr
&{}&\cr
\{ Q_\a, q_A\}&=&\g^i_{A\a} P_i  \ \ .
\label{qiv}
\end{eqnarray}

Recall that the compactification radius $R$ was related to the string
coupling as $R=l_s g_s$, and that the RR photon of the IIA superstring
is the KK photon from compactifying $x^{11}$ on $S^1$ of radius $R$ with
the RR charge corresponding to $\pe$.
Recall also that no perturbative string states carry RR charge charge and
hence have vanishing $\pe$. RR photon charge is only carried by the D0
branes. A single D0-brane carries one unit of RR charge and thus has
$\pe={1\over R}$. It fills out a whole supermultiplet of 256 states.
Since in eleven dimensions it is massless (graviton multiplet), in ten
dimensions it is BPS saturated, as we indeed saw. There also are KK states
with $\pe={N\over R}$, $N$ being an arbitrary integer. For $N\ne 1$ this
does not correspond to an elementary D0-brane. $N>1$ are bound states of
$N$ D0-branes, while $N<0$ corresponds to anti-D0-branes or bound states
thereof. As we take the total $\pe$ to infinity to reach the IMF limit,
only positive $\pe$ should appear, i.e. $N>0$. This means that M-theory
in the IMF should only contain D0-branes and their bound states. What has
happened to the anti-D0-branes and the perturbative string states ($N=0$)?
The answer is that these states get boosted to infinite energy and have
somehow implicitly been  integrated out. This means that the D0-brane
dynamics in the IMF should know in some subtle way that before going to
the IMF, there was more to
M-theory and type IIA superstrings then just D0-branes. This is much as
in field theory where the IMF vacuum is trivial, but still, in the end, 
the amplitudes and cross sections know about vacuum polarisation and all
the subtle effects of quantum field theory.
Moreover, M-theory should also contain membranes (i.e. 2-branes) and
5-branes. Where are they? We will see below that membranes can
effectively be described within the D0-brane quantum mechanics. The
5-brane on the other hand seems to be more subtle.

I can now state the BFSS conjecture: M-theory in the IMF is a theory in
which the only dynamical degrees of freedom are D0-branes each of which
carries a minimal quantum of $\pe=1/R$. This system is decribed by the 
effective action for $N$ D0
branes which is a particular $N\times N$ matrix quantum
mechanics, to be taken in the $N\to\infty$ limit.

\subsection{\it The matrix model Hamiltonian}

The effective action for a system of $N$ D0-branes was already given in
eq. (\ref{dxvii}). It is this action which is the starting point for the
matrix model description of M-theory in the IMF. For convenience, I
repeat it again:
\begin{eqnarray}
S&=& T_0\int {\rm d}t\, \tr \Big( 
{1\over 2 g_s} (D_0X^i)^2 - i\th^T D_0 \th\cr
&+&{c^2\over 4 g_s}([X^i,X^j])^2 + c \th^T \g^j [X_j,\th] \Big) \ .
\label{qv}
\end{eqnarray}
(Recall that $c=1/(2\pi\a')$ and $T_0=1/\sqrt{\a'}=1/l_s$.)
The indices $i=1, \ldots 9$ run over the nine {\it transverse}
directions, and the $\th$ are sixteen-component real spinors. The $X^i$
and $\th$ are all in the adjoint representation of the gauge group ${\rm
U}(N)$, i.e. they are hermitian $N\times N$ matrices. The covariant
derivative $D_0$ contains the gauge field $A_0$, but let us make the
gauge choice $A_0=0$ to simplify things.
Note that the first tirm in the action just reads 
$\int {\rm d}t {M\over 2} ({\rm d}X^i/{\rm d} t)^2$ 
where $M=T_0/g_s$ is the D0-brane mass. 
 We also rescale the fields as
$X^i=g_s^{1/3} Y^i$. This corresponds to the Weyl rescaling of the
metric, meaning that we now measure lengths with the eleven dimensional
supergravity metric $G$ rather than the ten dimensional string metric
$g$. One can also rescale the time accordingly: $t=g_s^{1/3} l_p \tilde\tau$
extracting also a factor of eleven dimensional Planck length $l_p$ to make
$\tilde\tau$ dimensionless. One has $l_p= g_s^{1/3} l_s$ so that  
\begin{equation}
t=g_s^{2/3} l_s \tilde\tau = {g_s l_s\over g_s^{1/3}}\tilde\tau 
= {R\over g_s^{1/3}}\tilde\tau
\label{qvi}
\end{equation}
This is the choice  made in ref. \cite{BFSS}. But a dimensionless $\tilde\tau$ gives
a dimensionless Hamiltonian. For the discussion of 
the spectrum of the Hamiltonian, to be
interpreted as energies, it is preferable to work with a Hamiltonnian that has the
dimension of an energy, as usual (i.e. inverse time or inverse length). So I define
instead %
\begin{equation}
t=g_s^{2/3} \tau = 
{T_0 R\over g_s^{1/3}}\tau
\label{qvia}
\end{equation}
Denoting then $\d/\d \tau$ simply by a dot, the action becomes
\begin{eqnarray}
S&=&T_0^2\int {\rm d}\tau\, \tr \Big( 
{1\over 2 R T_0^2} (\dot Y^i)^2 - i {1\over T_0}\th^T \dot \th\cr
&+&c^2 {R\over 4 }([Y^i,Y^j])^2 + c R \th^T \g^j [Y_j,\th] \Big) \ .
\label{qvii}
\end{eqnarray}
Defining the conjugate momenta of $Y^i$ and $\th$ as usual, $\Pi_i=\dot
Y^i/R$ and $\pi=-i T_0 \th^T$, one obtains the Hamiltonian
\begin{eqnarray}
H&=&R\, \tr \Big( {1\over 2 } \Pi_i^2
-{c^2 T_0^2\over 4 }([Y^i,Y^j])^2 \cr
&{}& \phantom{ R\, \tr \Big(   }
- c T_0^2\th^T \g^j [Y_j,\th] \Big)\cr
&\equiv& R\, \tilde H \ .
\label{qviii}
\end{eqnarray}
Recall that $-{1\over 4} R c^2 T_0^2 \tr ([Y^i,Y^j])^2$ is a non-negative
potential.
In the $R\to\infty$ limit of uncompactified M-theory, finite energy
states of $H$ correspond to states whose $\tilde H$ energy vanishes.
To be more precise, we are looking for states with
\begin{equation}
\tilde H \vert \P \ra = {\epsilon\over N} \vert \P \ra
\Leftrightarrow
H \vert \P \ra = {R\over N} \epsilon \vert \P \ra    
\label{qix}
\end{equation}
with finite $\epsilon$. But recall that for a system of $N$ D0-branes the
total $\pe$ momentum is
\begin{equation}
\pe = {N\over R}    
\label{qx}
\end{equation}
so that the energy $E$ takes the form
\begin{equation}
E = {\epsilon\over \pe}    
\label{qxi}
\end{equation}
Provided we can identify $\epsilon$ with ${1\over 2} p_\perp^2$  this
gives us the desired dispersion relation of {\it eleven} dimensions in
the IMF. I will return to this point later when discussing the spectrum
in more detail.

\subsection{\it Coordinate interpretation}

In section 2, I already touched upon the interpretation of the $N$ 
eigenvalues of the $X^i$ as the position vectors of the $N$ D0-branes.
Let me now elaborate this point a bit more.

The potential $V(Y)=-{1\over 4} R c^2 T_0^2 ([Y^i,Y^j])^2$ in the  
Hamiltonian is the familiar
Higgs potential, analogous to $([\f,\f^+])^2$ in ${\cal N}=2$ super YM in
four dimensions. In field theory, the supersymmetric vacua have
$[\f,\f^+]=0$. Here we have quantum mechanics, not quantum field theory
and the expectation values of the scalars $Y^i$ do not give
superselection sectors (i.e. distinct vacua) but are collective
coordinates with corresponding quantum wave functions: they are not
frozen at $V(Y)=0$. Still, $V(Y)$ has flat directions
(minima) $[Y^i,Y^j]=0$
along which the $Y^i$ can be simultaneaously diagonalized:
$
Y^i={\rm diag} (y_1^i, y_2^i, \ldots y_N^i )$ where $y_a^i$ gives the $i^{\rm
th}$ coordinate of the $a^{\rm th}$ D0-brane. More generally, if the
branes are far apart from each other, loosly speaking, the $Y^i$ are large
and non-commutativity would cost much energy. (This will be seen more
precisely below.) In this case commuting
$Y^i$ are a good approximation and the D0-brane positions  are rather
well defined. As they get closer, non-commuting configurations become
more important (strings stretching between different branes) and the
individual positions can no longer be well defined. Space is intrinsically
non-commutative with ordinary commutative space only emerging at long
distances. Yet one has the sull super-Galilean invariance in the
transverse directions. Translations e.g. are given by $Y^i\to Y^i + d^i
{\bf 1}$ where ${\bf 1}$ is the unit matrix. This does not affect the
kinetic terms nor the commutator terms in the Hamiltonian or action and
hence is an invariance of the theory. Similarly, a Galilean boost
$Y^i\to Y^i + v^i t {\bf 1}$ only
affects the center of mass momentum to be defined below, but not the
relative momenta, nor the interaction terms.

Consider configurations where the $N\times N$ matrices $Y^i$ take
block-diagonal forms with $n$ blocks of size $N_1, N_2, \ldots N_n$ and
$\sum_a N_a=N$. This corresponds to $n$ widely separated clusters of 
D0-branes. One can define the distance between cluster $a$ and cluster
$b$ as
\begin{equation}
r_{ab}=\left\vert {1\over N_a} \tr Y_a - {1\over N_b} \tr Y_b \right\vert
.
\label{qxii}
\end{equation}
As an example take $N=2$ and $N_1=1,\ N_2=1$ and 
$Y^i=\pmatrix{\a_i&\b_i\cr \b_i^*&\delta_i\cr}$. 
Then $r_{1,2}=\left( \sum_{i=1}^9 (\a_i-\delta_i)^2 \right)^{1/2}$ is
large if at least for one value of $i$, say $i_0$, 
$\vert \a_{i_0}-\delta_{i_0}\vert$ is large. One has 
$\vert \a_{i_0}-\delta_{i_0}\vert
\ge {1\over 3} r_{1,2}$. Then for any $j$ one has
\begin{eqnarray}
&-&{1\over 2}\tr ([Y^{i_0},Y^j])^2 
=4\, {\rm Im}(\b_i\b_j^*)^2 + \cr
&{}&\phantom{XX}
+\vert \b_j (\a_{i_0}-\delta_{i_0}) 
- \b_{i_0}  (\a_{j}-\delta_{j}) \vert^2 \ .
\label{qxiii}
\end{eqnarray}
Except for non-generic configurations\footnote{
Of course, if e.g. $Y^j=Y^{i_0}$ one has $[Y^{i_0},Y^j]=0$ no matter
how large $r_{1,2}$ is. But this is highly non-generic.
}
this will be of order 
\begin{equation}
\vert \b_j\vert^2 (\a_{i_0}-\delta_{i_0})^2
\ge {1\over 9}\vert \b_j\vert^2 r_{1,2}^2 \ .
\label{qxiiia}
\end{equation}
This generalizes to the general case of larger blocks: 
for generic configurations,
$\tr ([Y^i,Y^j])^2 $ is at least of the order of the modulus squared of the
off-block-diagonal elements times the minimum of the $r_{ab}^2$ times
some numerical constant. The bottom line is that 
for well seperated clusters of D0
branes - defined by large $r_{ab}$ - generic off-block diagonal elements must
be small or else they give rise to large potential energies $\sim
r_{ab}^2$. 

One might think that this harmonic oscillator type potential
would lead to a finite ground state energy, but this is not true due to
the supersymmetry of the system. I will now explain why such clusters of
D0-branes correspond to a collection of ``supergravitons"

\subsection{\it The spectrum of $H$ and the supergravitons}
 
Begin by considering the simplest case, namely $N=1$, a single D0-brane.
Then $\pe=N/R=1/R$ and 
\begin{equation}
H_{(N=1)}={R\over 2} \Pi_i^2
\equiv {R\over 2}  p_\perp^2 = {p_\perp^2\over 2 \pe} \ .
\label{qxiv}
\end{equation}
This is the eleven dimensional relativistically invariant relation
between energy and momentum of a massless particle when written in the
IMF. Comparing with eq. (\ref{qiii}) we see that the Hamiltonian has been
shifted by the constant $\pe$. Moreover, there are also the 16 $\th$'s that generate a
$2^{16/2}=256$ dimensional supermultiplet. Hence for $N=1$ the spectrum
of the Hamiltonian is that of an eleven dimensional massless
supermultiplet of 256 states containing up to spin two: this is exactly
the supergravity multiplet, and BFSS call it the supergraviton.

For $N>1$ one separates the center-of-mass coordinates and momenta from
the relative ones:
\begin{eqnarray}
Y^i=Y^i_{\rm rel} + Y^i_{\rm cm} {\bf 1} \ & , & 
\ Y^i_{\rm cm} ={1\over N} \tr Y^i \cr
\Pi_i=\Pi^{\rm rel} + {1\over N} P_i^{\rm cm} {\bf 1} \ &,& 
\ P_i^{\rm cm} = \tr \Pi_i 
\label{qxv}
\end{eqnarray}
with $\tr Y^i_{\rm rel}= \tr \Pi^{\rm rel} = 0$.
Plugging this into the Hamiltonian (\ref{qviii}) one gets
\begin{equation}
H=H_{\rm cm}+H_{\rm rel}
\label{qxvi}
\end{equation}
where 
\begin{equation}
H_{\rm cm}= {R\over 2N} (P_i^{\rm cm})^2 
= {1\over 2 \pe}(P_i^{\rm cm})^2
\label{qxvii}
\end{equation}
with the correct factor $ {R\over N}={1\over \pe}$ exactly as it should.
The relative part of the Hamiltonian 
$H_{\rm rel}(Y^i_{\rm rel},\Pi^{\rm rel})$ exactly looks like the
original Hamiltonian $H(Y^i,\P)$ given in (\ref{qviii}), 
except that now all matrices are
traceless, i.e. in ${\rm SU}(N)$. It has been shown  \cite{WITP,SETHI,PORR}
that $H_{\rm rel}$ has zero-energy (threshold) bound states. For these bound states
the total energy is just given by the center-of-mass energy
\begin{equation}
E=E_{\rm cm}= {R\over 2N} (p_\perp^{\rm cm})^2 
= {1\over 2 \pe}(p_\perp^{\rm cm})^2
\label{qxviii}
\end{equation}
Again this is a full supergravity multiplet of 256 states, i.e. a
supergraviton.
Thus for any $N$ the spectrum contains single supergraviton states. They
correspond to the bound state at threshold. 

To see that the matrix
Hamiltonian can also describe arbitrary many such supergravitons, consider
the block decomposition of $Y^i$ and $\Pi_i$ discussed in the previous
subsection. If these matrices are exactly block-diagonal, i.e. if the
off-block-diagonal elements are strictly vanishing,\footnote{
Of course, one should not forget that these off-block-diagonal
elements  are part of a quantum mechanical Hamiltonian acting on a
state in some Hilbert space, so the appropriate statement should be:
``if on a given state $\vert \P \ra$ the off-block-diagonal elements
of the $Y^i$ and $\Pi_i$ are vanishing then $H \vert \P \ra = \sum_a
H_a \vert \P \ra$ where each $H_a$ is a trace  over (operator-valued)
$N_a\times N_a$ matrices.
} 
the total Hamiltonian
splits into a sum of $n$ uncoupled Hamiltonians $H_a$, one for each block
of size $N_a$. Small, but  non-vanishing off-block-diagonal elements
would correspond to interaction terms between these Hamiltonians. 
The Hamiltonians $H_a$ of course have exactly the same form as the 
initial Hamiltonian $H$. Again for each block matrix $Y_a^i$  one can
separate the center-of-mass and relative coordinates and momenta, and
arrive at the conclusion that each of the block Hamiltonians $H_a$ 
contains a supergraviton in its spectrum.  
Hence the matrix model can describe several
supergravitons, too. Since we want to be able to describe an arbitrary
number of them, we must let $N$ go to infinity. Then we conclude that the
matrix model contains the full Fock space of supergravitons. Since this
Fock space is embedded into the larger D0-brane quantum mechanics, the theory
should be free of UV divergences. The next question is whether the
matrix model also has something to say about the interactions of these
supergravitons. These interaction come about via the off-block-diagonal
matrix elements. As such they look very non-local in the first place. We
will nevertheless see that the matrix model, at low
energies, correctly reproduces the (local) supergraviton 
interaction of supergravity.

\subsection{\it Low-energy supergraviton scattering}

A convenient way to study the scattering of two
supergravitons with low {\it transverse} velocities
is to compute the  effective action by expanding the matrix model
action around a corresponding classical configuration. I will present
the one-loop computation following \cite{BECKER} where further details can be found. 
For a scattering process with relative transverse
velocity $v$ and impact parameter $b$ one can e.g. expand the $X^i$ as
follows:
\begin{eqnarray}
X^8&=&{1\over 2} v t\, \s_3 + \sqrt{g_s}\, \delta X^8 ,\cr
X^9&=&{1\over 2} b\,  \s_3 + \sqrt{g_s}\, \delta X^9 ,\cr 
X^i&=& \sqrt{g_s}\, \delta X^i,\ i\ne
8,9
\label{qxix}
\end{eqnarray}
where $\s_3$ is the Pauli matrix and the $\delta X$ are the quantum
fluctuations around the given classical configuration.  It is easy to see that the
classical configuration ($\delta X=0$) indeed corresponds to the desired
scattering process in  a reference frame where the total 
{\it transverse}
center-of-mass momentum and position vanish ($\tr \s_3=0$). Indeed,  
the $2\times 2$ matrices are block-diagonal corresponding to two 
``clusters" of D0-branes with $N_1=N_2=1$.
According to the definition
(\ref{qxii}) for the distance
between the two supergravitons one indeed finds 
$r\equiv r_{1,2}= \sqrt{(vt)^2+b^2}$. This is appropriate for two particles
that do not interact in a first approximation
and have impact parameter $b$. The interaction
will manifest itself only through a phase shift.

Expanding the action then
yields a collection of massless and massive modes depending 
on $b$, $v$ and $t$. As shown in \cite{BECKER}
the bosonic fields, including the gauge field, yield 16 modes
with masses $m^2=r^2_{1,2}\equiv r^2$, two modes with $m^2=r^2+2v$ and two others
with $m^2=r^2-2v$, as well as ten massless modes. There are also eight real 
fermions with masses $m^2=r^2+v$ and eight real 
fermions with masses $m^2=r^2-v$, as well as the ghost fields: two complex bosons
with $m^2=r^2$ and one massless complex boson.
Collecting all the determinants from integrating the massive fields we get
\begin{equation} 
D_{\rm tot}=D_0^{-6} D_{2v}^{-1} D_{-2v}^{-1} D_v^4 D_{-v}^4
\label{qxixa} 
\end{equation}
where, now with a euclidean time $\t$ (and a euclidean velocity, still denoted
by $v$),
\begin{eqnarray} 
D_\a &=& \det (-\d_\t^2 +r^2+\a)\cr   
&=& \det (-\d_\t^2 + b^2+ v^2\t^2+\a ) \ .
\label{qxixb} 
\end{eqnarray}
Note that the sum of the exponents on the right hand side of (\ref{qxixa})
vanishes, thanks to supersymmetry. Hence $\log D_{\rm tot}$ is not affected by
an additive (and possibly divergent) ambiguity $\log D_\a \to \log D_\a + d$,
provided the constant $d$ does not depend on $\a$.
Thus the (euclidean) effective one-loop action is 
\begin{equation} 
S_{\rm eff} =S_0-\log D_{\rm tot}
\label{qxixc} 
\end{equation}
so that one identifies the one-loop effective potential $V_{\rm eff}$ as
\begin{eqnarray} 
-\log D_{\rm tot} &=& \int {\rm d}\t  V_{\rm eff}(r(\t))\cr
&\equiv&
\int {\rm d}\t  V_{\rm eff}(\sqrt{b^2+v^2\t^2}) \ .
\label{qxixd} 
\end{eqnarray}
As usual, to obtain the one-loop effective potential, 
all one needs to do is to compute the determinants $D_\a$.

To this end, consider the quantum mechanical Hamiltonian of a harmonic oscillator of
unit mass
\begin{equation} 
H_\o ={1\over 2} (P^2 +\o^2 Q^2) \ ,\ [Q,P]=i \ .
\label{qxixe} 
\end{equation}
It is well known that the matrix elements of the euclidean evolution operator for a
time interval $2s$ are
\begin{eqnarray} 
&{}&\la q'\vert \e^{-(2s) H_\o} \vert q \ra \equiv U(\o,2s,q',q) \cr
&{}&\cr
&{}& = \left( { \o\over 2\pi \sinh 2s\o}\right)^{1/2} \times \cr
&{}&\cr
&{}& \times \exp \left( -{\o\over 2} 
{(q^2+{q'}^2) \cosh 2s\o -2 q q'\over  \sinh 2s\o} \right) .\cr
&{}&
\label{qxixf} 
\end{eqnarray}
On the other hand one has
\begin{eqnarray} 
&{}&\log \det (2H_\o+\l) = \tr \log (2H_\o+\l)\cr 
&{}&\cr
&{}&\simeq -\tr \int_0^\infty {{\rm d} s\over s}\, \e^{-2sH_\o-s\l}\cr 
&{}&\cr 
&{}&=-\int_0^\infty {{\rm d} s\over s}\, \e^{-s\l} 
\int_{-\infty}^\infty {\rm d} q\, 
U(\o,2s,q,q)  \ .
\label{qxixg} 
\end{eqnarray}
All these integrals are divergent as $s\to 0$. What is finite and makes
sense are the
derivatives with respect to $\l$. This is what I mean by the ``$\simeq$"
sign. Said differently, when expanding in powers of $\l$, one has an equality
for all terms with non-zero powers of $\l$, while the $\l$-independent terms
differs by a divergent constant. As remarked above, these divergent constants
cancel when computing $\log D_{\rm tot}$ and thus do not affect the validity
of the present computation.
Inserting then the explicit expression (\ref{qxixf}) 
for $U(\o,2s,q,q)$ into (\ref{qxixg}) and performing
the gaussian integration over $q$, one gets
\begin{equation} 
-\log \det (2H_\o+\l) \simeq \int_0^\infty {{\rm d} s\over s} 
{\e^{-s\l}\over 2 \sinh s\o} \ .
\label{qxixh} 
\end{equation}

Now observe that, if we replace $q$ by $\t$ so that 
$P^2+\o^2 Q^2\to  -\d_\t^2 +\o^2 \t^2$, the determinants (\ref{qxixb})
we are interested in are
\begin{equation} 
D_\a= \det (2H_v +b^2+\a) 
\label{qxixi} 
\end{equation}
i.e. $\l=b^2+\a$, so that
\begin{eqnarray} 
-\log D_{\rm tot} &=& \int_0^\infty {{\rm d} s\over s}
{\e^{-s b^2}\over 2\sinh sv} \times \cr
&{}&\cr
&\times& \left( -6-2\cosh 2sv +8 \cosh sv \right) 
\label{qxixj} 
\end{eqnarray}
As promised, there is no divergence as $s\to 0$.
For large impact parameter $b$ only small $s$ contribute significantly
to the integral and
\begin{equation} 
-\log D_{\rm tot} \simeq -{v^3\over b^6} + {\cal O}\left( {v^5\over b^{10}} \right) \ .
\label{qxixk} 
\end{equation}
Equation (\ref{qxixd}) then gives 
$\int {\rm d}\t  V_{\rm eff}(\sqrt{b^2+v^2\t^2})
=-{v^3\over b^6} + {\cal O}\left( {v^5\over b^{10}}\right)$ which yields the effective
long-range potential
\begin{equation} 
V_{\rm eff}(r)= -{15\over 16} {v^4\over r^7} 
+ {\cal O}\left( {v^6\over r^{11}}\right) \ .
\label{qxixl} 
\end{equation}

What is this supposed to mean for the scattering of eleven dimensional
(super)gravitons? Scattering in the IMF should be described by a non-relativistically
looking time-independent potential at vanishing $\pe$-transfer.
Remarkably, the potential (\ref{qxixl}) exactly coincides
with the corresponding result from eleven dimensional supergravity.
In particular, the factor $1/r^7$ comes from the eleven
dimensional propagator of massless fields. It is the  time-independent {\it space}
propagator at vanishing longitudinal momentum ($\pe$) transfer, 
i.e. integrated over
$x^{11}$: in $d$ space-time dimensions such a propagator is
$\int {\rm d}^{d-1}p\, \delta(p_L) \e^{i p x} / p^2 \sim 1/ x^{d-4}$,
which for $d=11$ indeed gives $1/r^7$. The velocity dependence is
also correct and, maybe even more remarkably, also the numerical
factor. So the one-loop matrix model computation already gives the
full and correct answer. Higher loop corrections to the
$v^4$ term would ruin this
agreement. Luckily, at two loops there are none  \cite{BECKER},
and probably the one-loop result is the full answer. 
The present computation was done for $N_1=N_2=1$ but it is easy to
reinstate the dependence on arbitrary cluster sizes $N_1$, $N_2$.
%
%

It is remarkable that the simple matrix model  knows quite a lot
about propagating massless gravitons in eleven dimensions. This is
a  non-trivial check for the
matrix model description of M-theory in the IMF. But M-theory also
has membrane configurations. How can they appear in the matrix
model?

\subsection{\it Membranes in the matrix model}

In order to see how the matrix model could describe (super)membranes, let me
first discuss the description of the latter. Just as classical superstrings
can only exist in certain space-time dimensions, the classical supermembrane
also cannot exist for all $d$. But there does exist one in eleven dimensions.
It is described by bosonic coordinates $y^\m(p,q,\tau)$  and their
superpartners. Here $p\equiv \s^1$, $q\equiv \s^2$ and $\tau$ are the three
world-volume coordinates, and the $y^\m$ describe how the membrane is
embedded into the eleven dimensional target space. In a Hamiltonian
formalism, no explicit $\tau$ dependence appears, 
$y^\m(p,q,\tau)\to y^\m(p,q)$, and all
$\tau$-derivatives are replaced by the corresponding  momenta $\pi_\m(p,q)$.
In the light-cone frame only the transverse components with $i=1, \ldots 9$
are dynamical and the membrane Hamiltonian is  \cite{MEMBRANE}
\begin{eqnarray}
H_{\rm m}
&=& {1\over 2 \pe}
\int { {\rm d}p{\rm d}q\over (2\pi)^2}\, \pi_i^2(p,q)\cr
&{}&\cr
&+& {(2\pi T_2^{\rm m})^2\over 4 \pe}\int  {\rm d}p{\rm d}q
\left( \{y^i(p,q),y^j(p,q)\}\right)^2\cr
&{}&\cr
&+& \ {\rm fermion\ terms}
\label{qxxii}
\end{eqnarray}
where I used the suggestive (standard) notation
\begin{equation}
\{A,B\} = \d_q A \d_p B - \d_p A \d_q B
\label{qxxiii}
\end{equation}
for any two functions $A,B$ of $q$ and $p$ and where $T_2^{\rm m}$ is the
membrane tension as I will show soon. The analogy with the matrix model
Hamiltonian is striking. Basically all one needs to do is to trade the two
discrete matrix indices for the continuous variables $q,p$ in the limit where
the size $N$ of the matrices goes to infinity.  The membrane world-volume is
taken to factorize as $\Sigma \times {\bf R}$ with some Riemann surface
$\Sigma$. 
I will only deal with
membranes of toric topology, i.e. with $\Sigma$ being a torus. 
Now let me show that the second term in the membrane Hamiltonian is
correctly normalised. For a configuration with vanishing transverse
momentum, $\pi_i=0$, the mass ${\cal M}$ of the membrane is ${\cal M}^2=2\pe
H$. If furthermore there are no fermionic excitations, one has from eq.
(\ref{qxxii}) 
\begin{eqnarray}
{\cal M}^2&=&{(2\pi T_2^{\rm m})^2\over 2} \int {\rm d}p{\rm d}q
\left( \{y^i(p,q),y^j(p,q)\}\right)^2 \cr
&{}&
\label{qxxiiia}
\end{eqnarray}
But the area ${\cal A}$ of the membrane is given by
\begin{eqnarray}
{\cal A}^2&=& (2\pi)^2 \int {\rm d}p{\rm d}q \sum_{i<j}
\left( \{y^i(p,q),y^j(p,q)\}\right)^2 \cr
&=& {1\over 2}(2\pi)^2 \int {\rm d}p{\rm d}q 
\left( \{y^i(p,q),y^j(p,q)\}\right)^2 \cr
&{}&
\label{qxxiiib}
\end{eqnarray}
as one can see by considering the special case $y^8(p,q)={p\over 2\pi} L_8$, 
$y^9(p,q)={q\over 2\pi} L_9$, $p,q\in[0,2\pi]$ with ${\cal A}=L_8 L_9$.
Thus one sees that ${\cal M}^2=(T_2^{\rm m} {\cal A})^2$ so that 
$T_2^{\rm m}$ is indeed the membrane tension.

Let me now show how the matrix model can yield the above membrane
Hamiltonian and what its prediction for the membrane tension is. For toric
membranes, the  $y^i(p,q)$ are doubly periodic functions and their expansions
yield the Fourier modes $y^i_{mn}$. These form nine $\infty\times \infty$
matrices just as would do the nine $Y^i$ of the matrix model in the
$N\to\infty$ limit. However, one still needs to perform a change of basis, so
that the commutator $[Y^i,Y^j]$ directly goes over into the bracket
$\{y^i,y^j\}$. This is achieved by the following trick.

On the space of $N\times N$ matrices introduce two matrices $U$ and $V$ such
that 
\begin{equation}
U^N=V^N={\bf 1} \ {\rm and} \ UV=\e^{2\pi i/N} VU\ .
\label{qxxiv}
\end{equation}
A particular realisation is given by $U_{j,j+1}=U_{N,1}=1$ and $V_{j,j}=
\e^{2\pi i (j-1)/N}$ with all other matrix elements vanishing. A more
abstract realisation is
\begin{equation}
U=\e^{i p} \ ,\ V=\e^{i q} \ ,\ [q,p]={2\pi\over N} i
\label{qxxv}
\end{equation}
in terms of two operators/matrices that behave like position and momentum on
a discrete and compactified space: $U^N=V^N={\bf 1}$ implies that $p$ and $q$
have eigenvalues $0,\ {2\pi\over N},\ 2\, {2\pi\over N}, \ldots 
(N-1)\, {2\pi\over N}$.  It follows that
\begin{equation}
\tr U^n V^m = N\, \delta_{n, 0 \mod N}\, \delta_{m, 0 \mod N} \ .
\label{qxxvi}
\end{equation}
This allows us to ``Fourier" decompose any $N\times N$ matrix $Z$ as follows
\begin{eqnarray}
Z&=&\sum_{n,m=-{N/2}-1}^{N/2} z_{nm} U^n V^m \ , \cr
z_{nm}&=&{1\over N}
\tr U^{-n} Z V^{-m} \ .
\label{qxxvii}
\end{eqnarray}
If the matrices $U$ and $V$ are written as $\e^{i p}$ and $\e^{i q}$ then one
simply has
\begin{equation}
Z=\sum_{n,m=-{N/2}+1}^{N/2} z_{nm} \e^{i n p} \e^{i m q}
\label{qxxviii}
\end{equation}

Now consider what happens in the $N\to\infty$ limit. As $N\to\infty$, $p$ and
$q$ commute and their eigenvalues fill  $[0,2\pi]\times [0,2\pi]$ with $2\pi$
and 0 identified, i.e. $(p,q)$ take values on a two-torus. Equation
(\ref{qxxviii}) then really is nothing but the standard Fourier decomposition
of a double-periodic function on a circle. Let's call this function
\begin{equation}
z(p,q)=\sum_{n,m=-\infty}^\infty z_{nm} \e^{i n p} \e^{i m q}
\label{qxxix}
\end{equation}
with, of course,
\begin{equation}
z_{nm}=\int_0^{2\pi}\int_0^{2\pi}{{\rm d}p\over 2\pi}{{\rm d}q\over 2\pi}
z(p,q) \e^{-i n p} \e^{-i m q}.
\label{qxxx}
\end{equation}
Since $\tr Z = N z_{00}$ it follows that in the $N\to\infty$ limit one has
\begin{equation}
\tr Z \to N \int_0^{2\pi}\int_0^{2\pi}{{\rm d}p\over 2\pi}{{\rm d}q\over 2\pi}
z(p,q) \ .
\label{qxxxa}
\end{equation}
Next, I will show that the commutator of two matrices goes over to the
bracket (\ref{qxxiii}) of the two corresponding functions. First note that
$[U^n, V^m]
= 2i \sin {nm\pi\over N} \e^{inp+imq}$. It follows that
\begin{eqnarray}
&{}&{N\over 2\pi i} [U^nV^k, U^mV^l]\cr
&{}&\cr
&{}&= (nl - km) \e^{i(n+m)p+i(k+l)q} + {\cal O} (1/N) 
\label{qxxxi}
\end{eqnarray}
But as $N\to\infty$ this precisely goes to the bracket
$\{ u^n v^k, u^m v^l\}$ where
$u(p,q)=\e^{ip}$ and $v(p,q)=\e^{iq}$ are the classical functions associated
to $U$ and $V$. By the bilinearity of the commutator and the bracket it then
follows for any two $N\times N$ matrices $Z, W$ and their associated
classical functions $z(p,q), w(p,q)$ one has
\begin{equation}
{N\over 2\pi i} [Z,W] \to  \{z(p,q), w(p,q)\} \ .
\label{qxxxii}
\end{equation}
Now given this correspondence and the form of the matrix model and
supermembrane Hamiltonians it is clear that the latter will turn out to be
the large $N$ limit of the former. However, one has to carefully define the
conjugate momentum: the classical function $\pi_i(p,q)$ corresponding to the
matrix $\Pi_i$ is not the canonical conjugate momentum of $y^i(p,q)$
corresponding to $Y^i$, but they differ by a factor of $N$ as one can see by
first working out the large $N$ limit of the Lagrangian (\ref{qvii}) (I will
only write the bosonic terms):
\begin{eqnarray}
L_{\rm matrix}^{\rm bos} &\to& {N\over 2R} 
\int {{\rm d}p\over 2\pi}{{\rm d}q\over 2\pi}
(\dot y^i(p,q))^2 \cr
&-& {R\over 4 N} c^2 T_0^2 \int {{\rm d}p}{{\rm d}q} (\{y^i,y^j\})^2 
\label{qxxxiii}
\end{eqnarray}
Recall that $N/R =\pe$ so that the momentum conjugate to $y^i(p,q)$ is
$\pi_i(p,q)=\pe \dot y^i(p,q)$ (and not ${1\over R} \dot y^i$). It follows
that the $N\to\infty$ limit of the matrix model precisely gives the
Hamiltonian (\ref{qxxii}) of the supermembrane with a membrane tension given
by
\begin{equation}
(2\pi T_2^{\rm m})^2= c^2 T_0^2 = {1\over (2\pi \a')^2} T_0^2 \ .
\label{qxxxiiia}
\end{equation}
But according to eq. (\ref{dxiva}), $T_0/(2\pi\a')=2\pi T_2$ so that the
matrix model yields a membrane with tension $T_2^{\rm m}=T_2$ equal to the
D2-brane tension,\footnote{ 
Recall that the true D2-brane tension is
$\tau_2$, while  $T_2=\t_2 g_s$. 
Now the membrane tension in M-theory is the D2-brane
tenion $\tau_2$ times the factor $(g_s^{1/3})^3$ due to the Weyl rescaling,
since the M-theory tension should be measured with the eleven dimensional
supergravity metric, while the D2-brane tension was measured with the ten
dimensional string metric. This means that the M-theory membrane tension
must be $\tau_2 g_s=T_2$ as claimed.
}  
i.e. the correct membrane tension of M-theory!

We have seen how a given matrix configuration in the large $N$ limit yields
some  membrane configuration, although generally a highly irregular one. 
Conversly to obtain a
given membrane configuration of toroidal topology, one starts with its
embedding functions $y^i(p,q)$ in the light cone, computes its Fourier
coefficients $y^i_{mn}$ and defines for every finite $N$ the matrix 
$Y^i_{(N)}=\sum_{m,n=-N/2+1}^{N/2} y^i_{mn} U^m V^n$. In particular, if the
membrane is smooth, coefficients $y^i_{mn}$ with large $m$ or $n$ will be
small, and the information lost about the membrane by including only
$\vert m\vert ,\vert n\vert\le N/2$ will be small.
To describe membranes of non-toroidal topology in the matrix model is more
subtle. An exception is the plane membrane e.g. obtained from the example following
eq. (\ref{qxxiiib}) in the limit $L_8, L_9\to\infty$.

\section{CONCLUSIONS AND NO FURTHER DEVELOPMENTS}

I will not talk about further developments, not because there are
none but because there are too many. I will not even attempt to
give references. So let me just briefly conclude what we have seen.

In the first lecture, I briefly reviewed D-branes explaining why $N$
D$p$-branes should be described by ten dimensional ${\rm U}(N)$ super
Yang-Mills theory reduced to $p+1$ dimensions. In  the second lecture, I
introduced M-theory as the eleven dimensional theory that, when formulated on
${\bf R}^{10}\times S^1$ with $S^1$ of radius $R$, is equivalent to the
IIA superstring on ${\bf R}^{10}$ and with string coupling constant $g_s=R
\sqrt{\a'}$ In the third lecture, I developed the ideas of BFSS and described
their matrix model for M-theory in the infinite momentum frame, as well as
several checks of this conjecture: 1) the matrix model contains the full Fock
space of an arbitrary number of supergravitons (supergravity
multiplets of 256
states); 2) remarkably, it gives the correct result for low-energy
supergraviton scattering (including terms up to $\sim v^4$) up to and
including a matrix-model two-loop computation; 3) the matrix model contains
(super) membranes, and in the large $N$ limit the matrix model dynamics goes over to
the dynamics of the corresponding (super)membranes. The tension of these
matrix model membranes agrees with the tension of the M-theory membranes.

\vskip 3.mm
\noindent
{\bf Acknowledgements}

I am grateful to J.-P. Derendinger for the invitation to present these
lectures at the TMR conference ``Quantum aspects of gauge theories,
supersymmetry and unification" in Neuch\^atel. I also wish to thank all those
who attended these lectures in Neuch\^atel and Paris 
and, by their questions and remarks, improved this
written version. This work is partially
supported by the European Commision under TMR contract
ERBFMRX-CT96-0045.

\end{document}